\documentclass[review, authoryear]{elsarticle}

\usepackage[a4paper, left=4cm, right=4cm, top=4cm, bottom=4cm]{geometry}

\usepackage[english]{babel}
\usepackage[utf8]{inputenc}
\usepackage[T1]{fontenc}
\usepackage{color}
\usepackage{lmodern, microtype}

\usepackage{amsmath}
\usepackage{amstext}
\usepackage{amssymb}
\usepackage{latexsym}
\usepackage{dsfont}
\usepackage{mathrsfs}
\usepackage{setspace}
\usepackage{ulem}
\usepackage{bbold}

\usepackage{hyperref}
\hypersetup{
    colorlinks=true,
    unicode=true
}

\usepackage{graphicx}
\usepackage{float}
\usepackage{caption, booktabs}
\usepackage{graphics}
\usepackage{subcaption}
\usepackage{lipsum}
\usepackage{amsmath,amssymb,amsfonts,amsthm}
\usepackage[table, dvipsnames]{xcolor}
\usepackage{mathrsfs}
\usepackage{soul}

\usepackage{threeparttable}

\newcommand{\R}{{\mathbb R}}

\newcommand{\bsu}{\boldsymbol u}

\newcommand{\bsx}{\boldsymbol x}

\newcommand{\bsU}{\boldsymbol U}

\newcommand{\bsX}{\boldsymbol X}

\definecolor{forest}{HTML}{0003c2}
\newcommand{\SP}[1]{\color{black}{{#1}} \color{black}} 

\definecolor{lila}{HTML}{6e00c2}
\newcommand{\JB}[1]{\color{black}{{#1}} \color{black}}

\definecolor{cyan}{rgb}{0,0.5,.5}
\definecolor{dcyan}{rgb}{0,0.3,.3}

\definecolor{leaf}{HTML}{009e1d}
\newcommand{\AAr}[1]{\color{black}{{#1}} \color{black}} 

\usepackage{tikz}
\usepackage{lineno}
\modulolinenumbers[0]

\begin{document}


\begin{frontmatter}

    \journal{Finance Research Letters (status: accepted)}

    \title{Modeling Volatility and Dependence of European Carbon and Energy Prices}


    \author[1]{Jonathan Berrisch}
    \ead{jonathan.berrisch@uni-due.de}

    \author[2]{Sven Pappert\corref{cor1}}
    \ead{pappert@statistik.tu-dortmund.de}
    \cortext[cor1]{Corresponding author}

    \author[1]{Florian Ziel}
    \ead{florian.ziel@uni-due.de}

    \author[2,3]{Antonia Arsova}
    \ead{arsova@statistik.tu-dortmund.de}

    \address[1]{Chair of Environmental Economics, esp. Economics of Renewable Energy, University of Duisburg-Essen, Germany}
    \address[2]{Chair of Econometrics, Department of Statistics, TU Dortmund University, Germany}
    \address[3]{RWI -- Leibniz Institute for Economic Research, Germany}

    \begin{abstract}
        We study the prices of European Emission Allowances (EUA), whereby we analyze their uncertainty and dependencies on related energy prices (natural gas, coal, and oil). We propose a probabilistic multivariate conditional time series model with a VECM-Copula-GARCH structure which exploits key characteristics of the data. Data are normalized with respect to inflation and carbon emissions to allow for proper cross-series evaluation. The forecasting performance is evaluated in an extensive rolling-window forecasting study, covering eight years out-of-sample. We discuss our findings for both levels- and log-transformed data, focusing on time-varying correlations, and in view of the Russian invasion of Ukraine.
    \end{abstract}
    \begin{keyword}
        Carbon Prices \sep Conditional Volatility \sep Copula \sep Emission Allowances \sep Energy Markets \sep Forecasting \sep Multivariate Modeling \sep Time Series
        \JEL C21 \sep Q41 \sep Q5 \sep C32 \sep Q59 \sep C58 \sep G17
    \end{keyword}
\end{frontmatter}

\newpage
\section{Introduction}
The European Union (EU) emissions trading system (ETS) allows polluters to trade emission allowances (European Emission Allowances, EUA). Their supply is fixed, while the demand is determined by the market participants.
Understanding the underlying dynamics of this market is essential for policymakers, commodity trading, and sustainability planning. This motivates exploring the determinants of the EUA price and its statistical properties.

The EU-ETS was set up in 2005, making it the first ETS worldwide. Not much later, research picked up the market. The EU-ETS has significant implications for energy companies and the industrial sector. As a side-effect, the introduction of the EU-ETS has strengthened the link between gas and power prices, leading to undesirable geopolitical uncertainties in the power price. More than  fifteen years later, following the Russian invasion of Ukraine, the geopolitical uncertainties of natural gas supply are more pronounced than ever before. It is, therefore, plausible to assume that some dependence between energy assets and the EUA price exists. Early studies dealing with these relations support this view:
\citet{fezzi2009interaction} employ a cointegrated VAR model and find gas prices to induce quick reactions of EUA prices, while electricity prices show a delayed reaction to shocks in EUA prices.

The dependence of EU-ETS on other energy markets (including coal, oil, natural gas, and electricity) has  been analyzed more recently by \citet{chevallier2019conditional}, who find a negative link between carbon prices and oil and gas prices. \citet{meier2020commodities} also find that EUA prices are significantly influenced by fossil fuel prices, while \cite{zhou2020measurements} show evidence that the carbon price drives the price of renewable energies. 
Employing an artificial neural network,  ~\cite{garcia2020short} pronounce the effect of subjective economic and political decisions on the EUA price and reject the influence of the external variables considered (iron, electricity, and steel prices). 
\citet{duan2021marginal} find a negative impact of energy prices on EUA prices. The effect is asymmetric, i.e., it is more pronounced in the lower quantiles of the EUA distribution.

Some studies focus on forecasting EUA prices, which is particularly relevant for portfolio management. \citet{paolella2008econometric} look into forecastability and value-at-risk forecasting in the EU-ETS, employing an AR-GARCH model with generalized skewed student-t innovations. \citet{benz2009modeling} use a Markov-switching framework including GARCH effects to model EUA prices, while \citet{trabelsi2022co2} compare the GARCH approach to Generalized Autoregressive Score models.
Many of these studies find EUA prices to exhibit heavy tails, leverage effects, and asymmetries. Therefore student-t and skewed-student-t distributions are often used and have been shown to lead to superior performance in forecasting \citep{trabelsi2022co2, benz2009modeling, paolella2008econometric}.
%

Another strand of the literature deals with the volatility process of EUA prices and possible spillovers to other markets. \citet{dutta2019assessing} find structural breaks in the EUA volatility process. They show that the forecasting performance of GARCH models increases (while persistence decreases) when accounting for structural breaks. \citet{hanif2021nonlinear} investigate volatility spillovers between EUA prices and energy indices using copulas. Finally, \citet{reboredo2014volatility} analyzes volatility spillovers between oil and EUA markets. Their results suggest leverage effects and volatility dynamics but do not confirm significant spillovers.

The brief literature review above highlights several key characteristics of EUA prices which have implications for modeling and forecasting. First, quite a few studies point towards a relationship to the prices of related commodities like oil, gas, and coal. Second, a normality assumption is likely to fail due to the excess kurtosis and possible skewness of the data. Third, there is evidence for volatility dynamics and even structural breaks.
%

In this paper, we focus on probabilistic multi-step ahead forecasting of EUA prices. Building upon the aforementioned findings of earlier studies, we propose a VECM-Copula-GARCH model to jointly model EUA, gas, oil and coal prices as a four-variate system. As a combination of three different components the model is able to account for possible long-run equilibrium relations between carbon and fuel prices, autoregressive effects, conditional heteroscedasticity in the individual series as well as for \AAr{time-varying} dependence structures between them in a simultaneous yet flexible way. In contrast to earlier combination models which are estimated in stages, we utilize a one-step estimation approach to improve efficiency.

The contributions of this paper are manifold. First and foremost, we jointly model EUA, oil, coal, and natural gas prices using a probabilistic approach. The obtained multivariate predictive distribution can be used for sampling future trajectories and therefore enables probabilistic multi-step forecasts which exploit the dependencies of EUA on other commodity markets. \AAr{Second, we evaluate the proposed model in an extensive rolling-window forecasting study for both data in levels and after log-transformation, whose properties differ (e.g. variance, cointegration).} Several strictly proper scoring rules are employed to evaluate the multivariate forecasts as well as the marginal forecasts. Moreover, we test for significant differences between the forecasts \AAr{against competing standard benchmarks. Finally, we consider real price data, normalized by the carbon emissions of the respective commodity, to allow for scale-invariant evaluation.}

The remainder of this paper is organized as follows. The data are presented in Section~\ref{sec_data}. Section~\ref{sec_model} introduces the proposed VECM-Copula-GARCH model and discusses its one-step estimation by maximum likelihood (ML). The set-up for the forecasting study and forecast evaluation measures are outlined in Section~\ref{sec_evaluation}, while Section~\ref{sec_results} discusses the results. Finally, Section~\ref{sec_conclusion} concludes.

\section{Data description and preliminary analysis}\label{sec_data}

\begin{figure}[!h]
    \centering
    \resizebox{\textwidth}{!}{%
        \fbox{\input{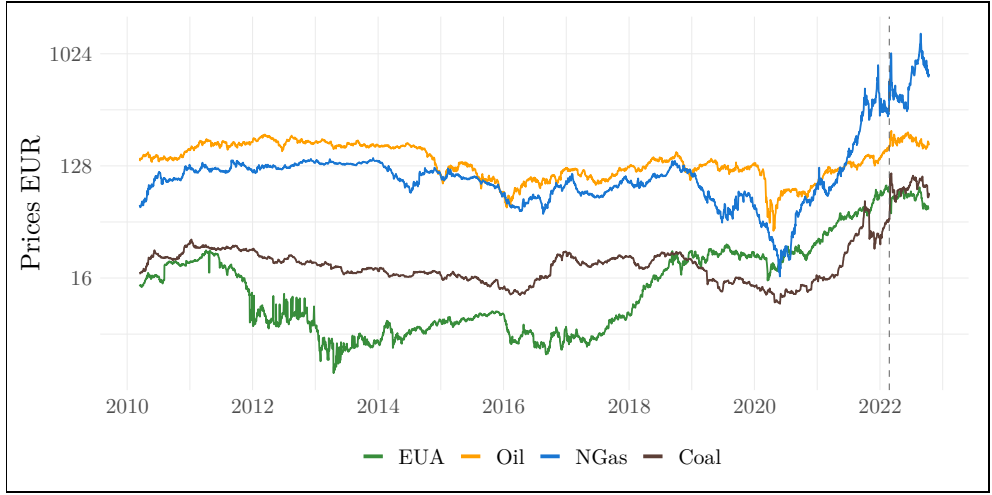}}
    }
    \caption{Time series plot of the data in levels. The dashed line marks the Russian invasion of Ukraine (2022-02-24).}
    \label{Fig_dataplot}
\end{figure}

The data, plotted in Figure \ref{Fig_dataplot}, comprise 3257 daily observations of short-term futures prices covering the period from March 15, 2010, until October 14, 2022, on four commodities, traded on the Intercontinental Exchange (ICE): European Allowances (carbon emissions), natural gas (Dutch TTF natural gas futures), coal (Rotterdam coal futures), and Brent crude oil. Oil and coal prices, originally given in USD, have been converted from USD into EUR. 
\AAr{All depicted price series have been standardized by the CO$_2$ emissions of the respective commodity (for conversion factors see \cite{quaschning2016understanding}). This means that the emission-adjusted prices reflect one tonne of CO$_2$ emissions for each of the three commodities. EUA prices needed no adjustment since they already reflect one tonne of CO$_2$ emissions. Furthermore, we adjusted for inflation by Eurostat's HICP, excluding energy\footnote{Source: \href{https://ec.europa.eu/eurostat/databrowser/bookmark/13937a8d-e341-406d-aacb-a6c34f93c175?lang=en}{Eurostat} (HICP - monthly data, Overall index excluding energy: \textit{TOT\_X\_NRG}, Euro Area: \textit{EA}, Base Year 2015: \textit{I15})}. Thus the general price increase over time is accounted for without dampening the commodities' price volatility which is to be explained by the model.}

Raw data have been transformed by taking natural logarithms following the literature convention  to stabilize their variance. Unit root testing by the augmented Dickey-Fuller (ADF) test points to non-stationarity in the levels \AAr{and logarithms of all four series, respectively,} and stationarity in their first differences\footnote{Results of unit root and cointegration tests are omitted for brevity but are available upon request.}. \AAr{As for cointegration, Johansen's likelihood ratio trace test points to two cointegrating relationships in the system in levels, which vanish in the log-transformed data due to the non-linear nature of the transformation. Subsequently, both data in levels and in logarithms have been considered for the estimation and evaluation of the proposed model.}
This allows us to evaluate whether accounting for long-term equilibrium relationships contributes to better forecasting performance.

\section{The VECM-Copula-GARCH model}\label{sec_model}
\subsection{Model design}

The preliminary analysis in the previous section reveals several key characteristics of the data which we explicitly account for in our proposed model. First, the variables are non-stationary \AAr{and cointegrated in the levels.}  While cointegration is not formally detected in the \SP{log-}transformed data, we wish to explore whether allowing for it may nevertheless improve forecasting performance, so we employ a VECM for the mean process, and experiment with different cointegrating ranks, including rank zero, i.e. no cointegration. Second, consistent with the literature, we consider univariate GARCH models for the conditional heteroscedasticity of each series. Third, we model the conditional cross-sectional dependence structure of the individual time series by a copula, thereby accounting for both linear and non-linear dependence in a consistent manner. For more flexibility, the dependence is allowed to vary over time, which is implemented by modeling the copula parameters by GARCH-type \SP{models as in e.g. \cite{jondeau2006copula}}. 

Denoting the $K$-dimensional vector of observable variables at time $t$ by $\bsX_t$, our general model can be formulated in terms of the conditional joint distribution \SP{ $F_{\bsX_t|\mathcal{F}_{t-1}}$, where $\mathcal{F}_{t}$ is the sigma field generated by all information available up to and including time $t$. Using Sklar's Theorem, we specify the joint conditional distribution $F_{\bsX_t|\mathcal{F}_{t-1}}$ of the model variables in terms of their marginal distributions
    $F_{X_{k,t}|\mathcal{F}_{t-1}}$ for $k=1,\ldots, K$, and the copula $C_{\bsU_{t}|\mathcal{F}_{t - 1}}$,} as is popular in many financial applications \citep{jondeau2006copula, hu2006dependence}. Thus, for the realized values $\bsx_t= (x_{1,t},\ldots, x_{K,t})^\intercal$ it holds  that \SP{
    \begin{align}
        F_{\bsX_t|\mathcal{F}_{t-1}}(\bsx_t)
        = C_{\bsU_{t}|\mathcal{F}_{t - 1}}(\bsu_t),
        \label{eq_F_marg_cop_decomp}
    \end{align}
}
where $\bsu_t=(u_{1,t},\ldots, u_{K,t})^\intercal$ and $u_{k,t} = F_{X_{k,t}|\mathcal{F}_{t-1}}(x_{k,t})$, $k=1, \ldots, K$.
That is, we specify $K$ marginal model components along with a copula that captures their dependency structure. In order to simplify the notation we drop the conditioning on $\mathcal{F}_{t-1}$ in equation~\eqref{eq_F_marg_cop_decomp} and write henceforth $F(\bsx_t) = C(\bsu_t)$.

The model can thus be specified as follows:
\SP{
    \begin{align}
        F(\bsx_t) = C\left[\mathbf{F}(\bsx_t;\mbox{\boldmath$\mu$}_t, \mbox{\boldmath$\sigma$}_{t}^2, \mbox{\boldmath$\nu$}, \mbox{\boldmath$\lambda$});\Xi_t, \Theta\right], \label{eq_copula}
    \end{align}
}
where the \SP{$(K(K-1)/2)$-dimensional} vector $\Xi_{t}$ comprises the time-varying dependence parameters, while the remaining time-invariant copula parameters are gathered in the vector $\Theta$. Specifically, we take $C$ as the $t$-copula to allow for heavy tails, in which case $\Xi_{t}$ stands for the time-varying correlation matrix, while $\Theta$ denotes the degrees of freedom parameter.
The individual marginal distributions, succinctly represented by the vector $\mathbf{F} = (F_1, \ldots, F_K)^{\intercal}$ in Eq. \eqref{eq_copula}, are specified in terms of the $(K\times 1)$ vectors of time-varying marginal means $\mbox{\boldmath$\mu$}_t$=$(\mu_{1,t}, \ldots, \mu_{K,t})^{\intercal}$ and variances $\mbox{\boldmath$\sigma$}^2_t$ $=(\sigma^2_{1,t}, \ldots, \sigma^2_{K,t})^{\intercal}$, respectively. In order to allow for both skewness and heavy tails in the marginals, we consider the generalized non-central $t$-distribution for each $F_i$, $i=1,\ldots, K$, with individual degrees of freedom and non-centrality parameters gathered in \mbox{\boldmath$\nu$} $= (\nu_1, \ldots, \nu_K)^{\intercal}$ and  \mbox{\boldmath$\lambda$} $= (\lambda_1, \ldots, \lambda_K)^{\intercal}$. For estimation, each $F_i$ has been parameterized so that its expectation and the variance are given by the parameters to be estimated.

We now turn our attention to the mean and variance processes $\mbox{\boldmath$\mu$}_t$ and $\mbox{\boldmath$\sigma$}^2_t$. The temporal evolution of the means vector is modeled by a VECM model as
\SP{\begin{align}
        \Delta \mbox{\boldmath$\mu$}_t = \Pi \bsx_{t-1} + \Gamma \Delta \bsx_{t-1},
        \label{eq_vecm}
    \end{align}}
where $\Pi = \alpha \beta^{\intercal}$, $\alpha \in \R^{K \times r}$ and $\beta \in \R^{K\times r}$ is the cointegrating matrix of rank $r$, $0 \leq r\leq K$.
\\
The dynamics of the individual variances are modeled by univariate GARCH models:
\begin{align*}
    \sigma_{i,t}^2 = & \omega_i + \alpha^+_{i} (\epsilon_{i,t-1}^+)^2 + \alpha^-_{i} (\epsilon_{i,t-1}^-)^2 + \beta_i \sigma_{i,t-1}^2,
\end{align*}
where $\epsilon_{i,t-1}^+ = \max\{\epsilon_{i,t-1}, 0\}$ and $\epsilon_{i,t-1}^- = \min\{\epsilon_{i,t-1}, 0\}$, $i=1, \ldots, K$, respectively. Separating the coefficients for positive and negative innovations allows the model to capture leverage effects.

Finally, the temporal evolution of the $K(K-1)/2$ primary dependence parameters of the copula, gathered in the (scaled) correlation matrix $\Xi_t$, is modeled similarly to a GARCH process:
\begin{align*}
    \Xi_{t} =    & \Lambda\left(\mbox{\boldmath$\xi$}_{t}\right),
    \\
    \xi_{ij,t} = & \eta_{0,ij} + \eta_{1,ij} \xi_{ij,t-1} + \eta_{2,ij} z_{i,t-1} z_{j,t-1},
\end{align*}
where $\xi_{ij,t}$ is a latent process, $z_{i,t}$ denotes the $i$-th standardized residual from time series $i$ \SP{at time point $t$}, and $\Lambda(\cdot)$ is a link function. The latter ensures that $\Xi_{t}$ does not exceed its support space and remains semi-positive definite by replacing negative values in the singular value decomposition of $\Xi_{t}$ by $10^{-5}$.

\subsection{Relevant nested models}

The proposed model \eqref{eq_copula} is the most general model incorporating cointegration, conditional heteroscedasticity, and time-varying dependence parameters. This model is denoted by \textbf{$\text{VECM}_{\text{lev}, \text{ncp}}^{r, \sigma_t, \rho_t}$}. \JB{Here the subscripts $\textbf{lev}$ and $\textbf{ncp}$ denote joint estimation along with the other model parameters of the leverage and non-centrality parameters, respectively,} while the superscripts signify
the assumed cointegrating rank ($r$), the time-varying conditional variance ($\sigma_t^2$) and the time-varying dependence parameters $(\rho_t)$.
The performance of this model will be compared to that of several nested models, following the same notational convention.  \textbf{$\text{RW}$}  will stand for a model with no cointegration such that the dynamics of the mean process can be thought of as a $K$-dimensional random walk. Assuming constant conditional variance and dependence parameters will be denoted by superscripts $\sigma^2$ and $\rho$, respectively, where the dependence on $t$ is suppressed. Hence, e.g., \textbf{$\text{VECM}^{r2, \sigma, \rho}$} will denote the VECM-Copula-GARCH model of cointegrating rank $r=2$, where the dependence parameters and variances are constant in time, and no leverage or non-centrality parameters are estimated. \AAr{Models estimated on log-transformed data are signified by the additional subscript $\text{log}$.}

\subsection{Estimation}
Despite having three different model components, all parameters can be estimated jointly by maximum likelihood (ML). Using conditional independence, the joint likelihood can be factorized as  \begin{align*}
    L = f_{X_1} \prod_{i=2}^T f_{X_i|\mathcal{F}_{i-1}},
\end{align*}
with the multivariate conditional density at any time point $t$ given by
\begin{align*}
    f_{\mathbf{X}_t}(\mathbf{x}_t | \mathcal{F}_{t-1}) = c\left[\mathbf{F}(\mathbf{x}_t;\mbox{\boldmath$\mu$}_t, \mbox{\boldmath$\sigma$}_{t}^2, \mbox{\boldmath$\nu$}, \mbox{\boldmath$\lambda$});\Xi_t, \Theta\right] \cdot \prod_{i=1}^K f_{X_{i,t}}(\mathbf{x}_t;\mbox{\boldmath$\mu$}_t, \mbox{\boldmath$\sigma$}_{t}^2, \mbox{\boldmath$\nu$}, \mbox{\boldmath$\lambda$}).
\end{align*}
Here the copula density $c$ can be derived analytically from the copula as $c(\mathbf{u}) = \partial_{\mathbf{u}} C(\mathbf{u})$.

\section{Forecasting study}\label{sec_evaluation}

\subsection{Study design}

To perform $h$-step ahead probabilistic forecasting from time point $t$ the following procedure is employed. First the model is fitted and the values of all time-varying parameters, $\Xi_t$, $\mbox{\boldmath$\mu$}_t$ and $\mbox{\boldmath$\sigma$}_{t}^2$ are determined using the information contained in $\mathcal{F}_{t}$. Then $n$ samples are drawn from the $K$-dimensional copula, $C\left(\cdot;\Xi_t, \Theta\right)$. The $i$-th marginal sample from the copula, $i=1,\hdots,K$, is quantile-transformed using the corresponding $i$-th quantile function,  $F_i^{-1}(\cdot;\mbox{\boldmath$\mu$}_t, \mbox{\boldmath$\sigma$}_{t}^2, \mbox{\boldmath$\nu$}, \mbox{\boldmath$\lambda$})$. The point forecast is estimated by the mean. Using the point forecast, all time-varying parameters are updated and the procedure is repeated $h-1$ times. Thus the generated samples will have a cross-sectional dependence structure imposed by the copula and marginal distributions according to the univariate densities. Expected values, variances, and other (mixed) moments can be approximated with arbitrary accuracy, which is limited only by computational power.

\subsection{Forecast evaluation}
Evaluation of the probabilistic forecast is conducted using different evaluation measures. The probabilistic multivariate forecasts will be evaluated by the energy score (ES). The energy score of a multivariate probabilistic forecast with distribution $F$ is given by
\begin{align*}
    \text{ES}_t(F, \mathbf{x}_t) = \mathbb{E}_{F} \left(||\tilde{\mathbf{X}}_t - \mathbf{x}_t||_2\right) - \frac{1}{2} \mathbb{E}_F \left(||\tilde{\mathbf{X}}_t - \tilde{\mathbf{X}}_t'||_2 \right),
\end{align*}
where $\mathbf{x}_t$ is the observed $K$-dimensional realization and $\tilde{\mathbf{X}}_t$, respectively $\tilde{\mathbf{X}}_t'$ are independent random vectors distributed according to $F$ \citep{gneiting2007strictly}. A low value of the ES indicates a good probabilistic forecast. In the univariate case, the energy score becomes the Continuous Ranked Probability Score (CRPS). \SP{Mean forecasts are evaluated by the root mean squared error (RMSE). Significant differences in forecast performances are tested with the Diebold-Mariano test \citep{diebold2002comparing} with the adjustment by \cite{harvey1997testing}.}

\section{Results}\label{sec_results}
\JB{We consider a rolling-window forecasting study with a window size of 1000 days which corresponds to about four years of data. As we have 3257 observations in total and compute 30-steps-ahead forecasts, this leaves 2227 potential starting points for the rolling-window study. To reduce computational costs, we sample $n=250$ data points uniformly from those 2227 grid points for further evaluation. We report the multivariate distributional forecast of all considered models by providing $2^{12}= 2048$ samples from the predicted distribution.}

We evaluate the full spectrum of the aforementioned VECM and RW models in conjunction with a \AAr{simple} exponential smoothing model, denoted by $\text{ETS}^{\sigma}$, \AAr{and a vector exponential smoothing model \SP{\citep{svetunkov2023new}}, denoted by $\text{VES}^{\sigma}$, } which we include as benchmarks. \SP{We choose a univariate and a multivariate ETS model as benchmarks since they are commonly used for forecasting \citep{jonsson2014exponential, abdul2019comparison, berrisch2022distributional}.}
For the sake of brevity, however, we report the performance only of selected ones, particularly the best-performing models.
\AAr{The performance of the models estimated on log-transformed data has been evaluated by taking the exponent of the forecasts without bias correction, in accordance with the recommendation of \cite{Demetrescu2020} for persistent data.} \JB{However, the exponent of some log forecasts evaluates to infinity. We replace these values with a large real integer ($10^5$) to enable a valid evaluation of these forecasts. This issue affects less than 0.001\textperthousand\ of all forecasts of the reported log models. We used $10^5$ as a replacement because it can be considered an extreme scenario for the prices in question. Note that we deal with inflation-adjusted, emission-normalized prices here. Evaluating the research on the social costs of carbon confirms that $10^5$ would be an extreme scenario, at least for EUA \citep{anthoff2013uncertainty, tol2019social, aldy2021keep}. The latter also holds for the other commodities if other price components are negligible.}

The performance of the selected models with respect to the ES can be found in Table \ref{tab:energy}. Their ES is compared to the ES of a simple random walk model with constant copula and no conditional heteroscedasticity modeling, $\text{RW}^{\sigma, \rho}$, \AAr{estimated on the data in levels.}
$\text{ES}_{1-30}^{\text{All}}$ denotes the score of the $(4 \times 30)$-dimensional multivariate ($4$-dimensional) 1-30 days ahead forecast. The general form of the ES allows us to evaluate the effects of modeling both the temporal and the cross-sectional dependence.
$\text{ES}_{1-30}^{\text{EUA}}$ denotes the score of the univariate 30-dimensional 1-30 days ahead forecast for the EUA prices (analogous for oil, natural gas, and coal prices). In this formulation of the ES, only temporal modeling of the time series is evaluated.
$\text{ES}_1^{\text{All}}$ denotes the score of the 4-dimensional multivariate one day-ahead forecast (analogously $\text{ES}_5^{\text{All}}$ denotes the score of the multivariate 5 day ahead forecast). This form of the ES evaluates the effect of modeling the cross-sectional dependence.
\begin{table}[!h]
    \caption{Improvement in energy score of selected models relative to $\textrm{RW}^{\sigma, \rho}_{}$ in \% (higher = better). Colored according to the test statistic of a DM-Test comparing to $\textrm{RW}^{\sigma, \rho}_{}$  (greener means lower test statistic i.e., better performance compared to $\textrm{RW}^{\sigma, \rho}_{}$). \label{tab:energy}}
    \centering
    \resizebox{\linewidth}{!}{
        \begin{tabular}{l>{}r>{}r>{}r>{}r>{}r>{}r>{}r>{}r}
            \toprule
            Model                                                                        & $\text{ES}^{\text{All}}_{1-30}$          & $\text{ES}^{\text{EUA}}_{1-30}$         & $\text{ES}^{\text{Oil}}_{1-30}$         & $\text{ES}^{\text{NGas}}_{1-30}$         & $\text{ES}^{\text{Coal}}_{1-30}$        & $\text{ES}^{\text{All}}_{1}$             & $\text{ES}^{\text{All}}_{5}$             & $\text{ES}^{\text{All}}_{30}$            \\
            \midrule
            $\textrm{RW}^{\sigma, \rho}_{}$                                              & \cellcolor[HTML]{BDBDBD}{161.96}         & \cellcolor[HTML]{BDBDBD}{10.06}         & \cellcolor[HTML]{BDBDBD}{37.94}         & \cellcolor[HTML]{BDBDBD}{146.73}         & \cellcolor[HTML]{BDBDBD}{13.22}         & \cellcolor[HTML]{BDBDBD}{5.56}           & \cellcolor[HTML]{BDBDBD}{13.28}          & \cellcolor[HTML]{BDBDBD}{34.29}          \\
            $\textrm{RW}^{\sigma_t, \rho_t}_{}$                                          & \cellcolor[HTML]{88C567}{9.40}           & \cellcolor[HTML]{93C966}{3.75}          & \cellcolor[HTML]{FFD346}{-0.41}         & \cellcolor[HTML]{84C467}{11.39}          & \cellcolor[HTML]{84C467}{4.13}          & \cellcolor[HTML]{A2CE64}{10.34}          & \cellcolor[HTML]{9BCC65}{9.10}           & \cellcolor[HTML]{A5CF64}{7.59}           \\
            $\textrm{RW}^{\sigma, \rho_t}_{\textrm{ncp}, \textrm{log}}$                  & \cellcolor[HTML]{6DBC69}{12.04}          & \cellcolor[HTML]{7FC268}{6.16}          & \cellcolor[HTML]{FFD849}{-0.56}         & \cellcolor[HTML]{6ABB6A}{14.33}          & \cellcolor[HTML]{66BA6A}{7.35}          & \cellcolor[HTML]{8DC766}{9.22}           & \cellcolor[HTML]{94C966}{9.82}           & \cellcolor[HTML]{90C866}{10.02}          \\
            $\textrm{RW}^{\sigma, \rho}_{\textrm{log}}$                                  & \cellcolor[HTML]{6DBC69}{12.10}          & \cellcolor[HTML]{7BC168}{6.25}          & \cellcolor[HTML]{FFD245}{-0.59}         & \cellcolor[HTML]{6ABB6A}{14.44}          & \cellcolor[HTML]{66BA6A}{7.31}          & \cellcolor[HTML]{8EC766}{9.04}           & \cellcolor[HTML]{93C966}{9.66}           & \cellcolor[HTML]{91C866}{9.91}           \\
            $\textrm{VECM}^{\textrm{r0}, \sigma_t, \rho_t}_{\textrm{lev}, \textrm{ncp}}$ & \cellcolor[HTML]{82C368}{9.68}           & \cellcolor[HTML]{FFC53C}{-0.72}         & \cellcolor[HTML]{E6E55C}{\textbf{0.32}} & \cellcolor[HTML]{7FC268}{11.74}          & \cellcolor[HTML]{88C567}{3.70}          & \cellcolor[HTML]{9BCC65}{10.82}          & \cellcolor[HTML]{9CCC65}{10.50}          & \cellcolor[HTML]{99CB65}{8.21}           \\
            $\textrm{VECM}^{\textrm{r0}, \sigma, \rho_t}_{\textrm{log}}$                 & \cellcolor[HTML]{6DBC69}{\textbf{12.15}} & \cellcolor[HTML]{71BD69}{6.10}          & \cellcolor[HTML]{FFCD42}{-0.70}         & \cellcolor[HTML]{6ABB6A}{\textbf{14.57}} & \cellcolor[HTML]{66BA6A}{7.80}          & \cellcolor[HTML]{93C966}{8.05}           & \cellcolor[HTML]{98CB65}{9.99}           & \cellcolor[HTML]{92C966}{\textbf{10.04}} \\
            $\textrm{VECM}^{\textrm{r1}, \sigma_t, \rho}_{\textrm{lev}, \textrm{ncp}}$   & \cellcolor[HTML]{75BF69}{10.65}          & \cellcolor[HTML]{FB9336}{-24.09}        & \cellcolor[HTML]{FFBA34}{-4.81}         & \cellcolor[HTML]{70BD69}{13.37}          & \cellcolor[HTML]{FFE351}{-0.93}         & \cellcolor[HTML]{A1CE64}{9.86}           & \cellcolor[HTML]{99CB65}{10.07}          & \cellcolor[HTML]{A6CF64}{7.10}           \\
            $\textrm{VECM}^{\textrm{r1}, \sigma, \rho_t}_{\textrm{ncp}, \textrm{log}}$   & \cellcolor[HTML]{9CCC65}{8.38}           & \cellcolor[HTML]{FFE855}{-0.40}         & \cellcolor[HTML]{FB8F38}{-4.65}         & \cellcolor[HTML]{92C966}{11.44}          & \cellcolor[HTML]{96CA66}{\textbf{8.66}} & \cellcolor[HTML]{A8D064}{6.62}           & \cellcolor[HTML]{B3D463}{7.98}           & \cellcolor[HTML]{F5EA5A}{1.42}           \\
            $\textrm{VECM}^{\textrm{r2}, \sigma_t, \rho}_{}$                             & \cellcolor[HTML]{97CB65}{8.99}           & \cellcolor[HTML]{FFA626}{-13.79}        & \cellcolor[HTML]{FFA527}{-7.15}         & \cellcolor[HTML]{8EC766}{11.43}          & \cellcolor[HTML]{FFE855}{-0.51}         & \cellcolor[HTML]{AED263}{9.86}           & \cellcolor[HTML]{A3CF64}{10.87}          & \cellcolor[HTML]{DCE15D}{4.32}           \\
            $\textrm{VECM}^{\textrm{r2}, \sigma, \rho_t}_{\textrm{log}}$                 & \cellcolor[HTML]{8AC667}{8.93}           & \cellcolor[HTML]{FFD849}{-2.77}         & \cellcolor[HTML]{FB9236}{-8.07}         & \cellcolor[HTML]{7BC168}{12.44}          & \cellcolor[HTML]{92C966}{6.58}          & \cellcolor[HTML]{9CCC65}{7.09}           & \cellcolor[HTML]{A0CE64}{9.42}           & \cellcolor[HTML]{D0DD5F}{4.42}           \\
            $\textrm{VECM}^{\textrm{r3}, \sigma_t, \rho_t}_{\textrm{lev}, \textrm{ncp}}$ & \cellcolor[HTML]{8FC766}{9.12}           & \cellcolor[HTML]{FFB330}{-9.18}         & \cellcolor[HTML]{FFBF38}{-3.68}         & \cellcolor[HTML]{88C567}{11.70}          & \cellcolor[HTML]{FFDD4D}{-1.01}         & \cellcolor[HTML]{AED263}{\textbf{10.94}} & \cellcolor[HTML]{ABD163}{\textbf{11.20}} & \cellcolor[HTML]{B5D562}{6.46}           \\
            $\textrm{VECM}^{\textrm{r3}, \sigma_t, \rho}_{\textrm{log}}$                 & \cellcolor[HTML]{FFED58}{0.05}           & \cellcolor[HTML]{FFB934}{-7.70}         & \cellcolor[HTML]{FFA928}{-6.51}         & \cellcolor[HTML]{F2E95A}{2.43}           & \cellcolor[HTML]{CEDD5F}{4.31}          & \cellcolor[HTML]{BBD662}{5.09}           & \cellcolor[HTML]{D6DF5E}{5.40}           & \cellcolor[HTML]{FFC039}{-17.97}         \\
            $\textrm{VECM}^{\textrm{r4}, \sigma_t, \rho}_{\textrm{lev}}$                 & \cellcolor[HTML]{E2E35D}{2.52}           & \cellcolor[HTML]{F88040}{-30.98}        & \cellcolor[HTML]{FB8F38}{-9.81}         & \cellcolor[HTML]{C3D961}{5.87}           & \cellcolor[HTML]{FFDD4D}{-2.04}         & \cellcolor[HTML]{AFD263}{9.00}           & \cellcolor[HTML]{BCD762}{7.29}           & \cellcolor[HTML]{FFBE37}{-9.23}          \\
            $\textrm{VECM}^{\textrm{r4}, \sigma, \rho_t}_{\textrm{ncp}, \textrm{log}}$   & \cellcolor[HTML]{CBDC60}{6.21}           & \cellcolor[HTML]{FFBB35}{-7.83}         & \cellcolor[HTML]{F46F47}{-15.69}        & \cellcolor[HTML]{B1D363}{10.43}          & \cellcolor[HTML]{B4D462}{6.83}          & \cellcolor[HTML]{BAD662}{5.54}           & \cellcolor[HTML]{C3D961}{7.28}           & \cellcolor[HTML]{FFDB4C}{-5.12}          \\
            $\textrm{ETS}^{\sigma}$                                                      & \cellcolor[HTML]{78C069}{9.94}           & \cellcolor[HTML]{88C567}{5.75}          & \cellcolor[HTML]{F6EA5A}{0.08}          & \cellcolor[HTML]{6EBC69}{13.05}          & \cellcolor[HTML]{8CC767}{7.83}          & \cellcolor[HTML]{ABD163}{6.96}           & \cellcolor[HTML]{A2CE64}{7.74}           & \cellcolor[HTML]{AED263}{6.21}           \\
            $\textrm{ETS}^{\sigma}_{\textrm{log}}$                                       & \cellcolor[HTML]{92C966}{8.12}           & \cellcolor[HTML]{85C467}{\textbf{7.80}} & \cellcolor[HTML]{FFD84A}{-0.51}         & \cellcolor[HTML]{87C567}{11.17}          & \cellcolor[HTML]{94C966}{8.54}          & \cellcolor[HTML]{CDDC60}{5.05}           & \cellcolor[HTML]{C9DB60}{6.14}           & \cellcolor[HTML]{E7E55C}{2.66}           \\
            $\textrm{VES}^{\sigma}$                                                      & \cellcolor[HTML]{A7D064}{5.50}           & \cellcolor[HTML]{FFB531}{-4.43}         & \cellcolor[HTML]{FFB330}{-3.22}         & \cellcolor[HTML]{A5CF64}{6.29}           & \cellcolor[HTML]{BCD762}{4.68}          & \cellcolor[HTML]{FB8F38}{-25.99}         & \cellcolor[HTML]{FFDC4C}{-2.42}          & \cellcolor[HTML]{D7E05E}{3.07}           \\
            $\textrm{VES}^{\sigma}_{\textrm{log}}$                                       & \cellcolor[HTML]{97CB65}{7.68}           & \cellcolor[HTML]{C0D861}{3.31}          & \cellcolor[HTML]{FFA329}{-4.34}         & \cellcolor[HTML]{95CA66}{9.07}           & \cellcolor[HTML]{85C467}{8.30}          & \cellcolor[HTML]{FC9733}{-22.11}         & \cellcolor[HTML]{F6EA5A}{1.07}           & \cellcolor[HTML]{D2DE5F}{4.32}           \\
            \bottomrule
            \multicolumn{9}{l}{\rule{0pt}{1em}Coloring w.r.t. test statistic: \colorbox[HTML]{66BA6A}{<-5} \colorbox[HTML]{7CC168}{-4} \colorbox[HTML]{91C866}{-3} \colorbox[HTML]{B0D363}{-2} \colorbox[HTML]{D8E05E}{-1} \colorbox[HTML]{FFED58}{0} \colorbox[HTML]{FFD145}{1} \colorbox[HTML]{FFB531}{2} \colorbox[HTML]{FC9733}{3} \colorbox[HTML]{F67744}{4} \colorbox[HTML]{EE5250}{>5} }                                               \\
        \end{tabular}}
\end{table}

The results of Table \ref{tab:energy} can be summarized as follows. \AAr{No model dominates uniformly in terms of forecasting performance for all variables and all horizons. Models estimated on the log-transformed data generally perform better than their counterparts estimated on the data in levels, despite the na{\"\i}ve forecasts of the former being potentially biased. This result can be explained by the log-transformation successfully stabilizing the variance of the series} \citep{Lutkepohl2012}. \AAr{As no cointegration was found in the log-transformed data, the VECM models are over-parameterized and are hence not expected to deliver superior forecasts. Indeed, considering all variables and all forecast horizons altogether, the best-performing models yield an improvement of ca. 12\% in $\text{ES}^{\text{All}}_{1-30}$ over the simple RW model in levels. This is achieved by the
$\text{VECM}^{r0}$
and the RW models with constant volatility and log transformation.
Allowing for time-varying dependence in the copula, or modeling the non-centrality parameter does not seem to influence their results much. These models also deliver the best forecasts at long horizons, and for natural gas prices in particular.}
\AAr{VECM models in levels provide slightly worse overall forecasts than the RW in logarithms but seem to perform best of all at short horizons with $r=3$.  Nevertheless, accounting for cointegrating relations offers at most modest improvement over the corresponding RW models in levels. Rather, allowing for time variation in the conditional variance and dependence parameters in the levels-RW models improves the forecast significantly and comparably to considering a simple RW in logarithms for $H=1,5$. The contribution of the leverage effects and the non-centrality parameter for the models in levels is subordinate. The forecasts of the level VECMs, however, seem quite unsatisfactory for the EUA and oil price series. They are rather best forecast by RW and ETS: the EUA price in logarithms, while the oil price in levels.}

Table \ref{tab:crps} presents the CRPS of the selected models for every univariate time series. H1, H5, and H30 denote the score \AAr{or improvement thereof} for the 1-, 5- and 30-day-ahead probabilistic forecast, respectively. The CRPS gives information about the quality of the univariate modeling only, as modeling the cross-sectional dependence has no effect on the univariate density obtained by quantile transforming the samples generated from the copula. Even though the samples from the copula have a dependence structure imposed by the copula, the marginal distributions are distributed uniformly.

\begin{table}[!h]
    \caption{Improvement in CRPS of selected models relative to $\textrm{RW}^{\sigma, \rho}_{}$ in \% (higher = better). Colored according to the test statistic of a DM-Test comparing to $\textrm{RW}^{\sigma, \rho}_{}$  (greener means lower test statistic i.e., better performance compared to $\textrm{RW}^{\sigma, \rho}_{}$). \label{tab:crps}}
    \centering
    \resizebox{\textwidth}{!}{
        \begin{tabular}[t]{l>{}r>{}r>{}r>{}r>{}r>{}r>{}r>{}r>{}r>{}r>{}r>{}r}
            \toprule
            \multicolumn{1}{c}{ }                                                        & \multicolumn{3}{c}{EUA}                & \multicolumn{3}{c}{Oil}                & \multicolumn{3}{c}{NGas}               & \multicolumn{3}{c}{Coal}                                                                                                                                                                                                                                                                                                                                                           \\
            \cmidrule(l{3pt}r{3pt}){2-4} \cmidrule(l{3pt}r{3pt}){5-7} \cmidrule(l{3pt}r{3pt}){8-10} \cmidrule(l{3pt}r{3pt}){11-13}
            Model                                                                        & H1                                     & H5                                     & H30                                    & H1                                     & H5                                     & H30                                    & H1                                      & H5                                      & H30                                     & H1                                      & H5                                     & H30                                    \\
            \midrule
            $\textrm{RW}^{\sigma, \rho}_{}$                                              & \cellcolor[HTML]{BDBDBD}{0.4}          & \cellcolor[HTML]{BDBDBD}{0.9}          & \cellcolor[HTML]{BDBDBD}{2.1}          & \cellcolor[HTML]{BDBDBD}{1.5}          & \cellcolor[HTML]{BDBDBD}{3.4}          & \cellcolor[HTML]{BDBDBD}{\textbf{9.1}} & \cellcolor[HTML]{BDBDBD}{4.7}           & \cellcolor[HTML]{BDBDBD}{11.6}          & \cellcolor[HTML]{BDBDBD}{29.8}          & \cellcolor[HTML]{BDBDBD}{0.3}           & \cellcolor[HTML]{BDBDBD}{0.9}          & \cellcolor[HTML]{BDBDBD}{2.8}          \\
            $\textrm{RW}^{\sigma_t, \rho_t}_{}$                                          & \cellcolor[HTML]{96CA66}{\textbf{5.6}} & \cellcolor[HTML]{90C866}{6.0}          & \cellcolor[HTML]{C8DB60}{2.8}          & \cellcolor[HTML]{BED761}{2.1}          & \cellcolor[HTML]{98CB65}{2.7}          & \cellcolor[HTML]{FFC73E}{-0.8}         & \cellcolor[HTML]{A3CF64}{12.6}          & \cellcolor[HTML]{A2CE64}{10.5}          & \cellcolor[HTML]{A1CE64}{9.6}           & \cellcolor[HTML]{88C567}{10.7}          & \cellcolor[HTML]{94CA66}{6.5}          & \cellcolor[HTML]{B3D463}{2.1}          \\
            $\textrm{RW}^{\sigma, \rho_t}_{\textrm{ncp}, \textrm{log}}$                  & \cellcolor[HTML]{9CCC65}{5.1}          & \cellcolor[HTML]{7AC068}{8.7}          & \cellcolor[HTML]{A9D064}{5.0}          & \cellcolor[HTML]{E3E45D}{0.7}          & \cellcolor[HTML]{E3E45C}{0.8}          & \cellcolor[HTML]{FFE04F}{-0.4}         & \cellcolor[HTML]{8EC766}{11.4}          & \cellcolor[HTML]{96CA66}{11.5}          & \cellcolor[HTML]{8FC866}{12.4}          & \cellcolor[HTML]{77BF69}{8.0}           & \cellcolor[HTML]{7FC268}{7.3}          & \cellcolor[HTML]{7CC168}{6.7}          \\
            $\textrm{RW}^{\sigma, \rho}_{\textrm{log}}$                                  & \cellcolor[HTML]{A3CF64}{4.7}          & \cellcolor[HTML]{76BF69}{8.9}          & \cellcolor[HTML]{A2CE64}{5.2}          & \cellcolor[HTML]{FFEC57}{0.0}          & \cellcolor[HTML]{F3E95A}{0.3}          & \cellcolor[HTML]{FFD84A}{-0.6}         & \cellcolor[HTML]{8EC766}{11.2}          & \cellcolor[HTML]{95CA66}{11.4}          & \cellcolor[HTML]{8FC866}{12.4}          & \cellcolor[HTML]{77C069}{7.7}           & \cellcolor[HTML]{7FC268}{7.5}          & \cellcolor[HTML]{7BC168}{6.6}          \\
            $\textrm{VECM}^{\textrm{r0}, \sigma_t, \rho_t}_{\textrm{lev}, \textrm{ncp}}$ & \cellcolor[HTML]{9BCC65}{3.6}          & \cellcolor[HTML]{E7E55C}{0.6}          & \cellcolor[HTML]{FFA928}{-1.6}         & \cellcolor[HTML]{B1D363}{\textbf{2.7}} & \cellcolor[HTML]{9DCD65}{\textbf{3.0}} & \cellcolor[HTML]{FFED58}{0.0}          & \cellcolor[HTML]{9CCC65}{13.1}          & \cellcolor[HTML]{A1CE64}{12.2}          & \cellcolor[HTML]{98CB65}{10.4}          & \cellcolor[HTML]{8EC766}{\textbf{11.8}} & \cellcolor[HTML]{92C966}{7.2}          & \cellcolor[HTML]{C5DA60}{1.5}          \\
            $\textrm{VECM}^{\textrm{r0}, \sigma, \rho_t}_{\textrm{log}}$                 & \cellcolor[HTML]{B2D363}{4.2}          & \cellcolor[HTML]{72BE69}{\textbf{8.9}} & \cellcolor[HTML]{99CB65}{5.1}          & \cellcolor[HTML]{FAEB59}{0.2}          & \cellcolor[HTML]{EFE85B}{0.4}          & \cellcolor[HTML]{FFD044}{-0.8}         & \cellcolor[HTML]{94C966}{9.9}           & \cellcolor[HTML]{9ACC65}{11.8}          & \cellcolor[HTML]{90C866}{\textbf{12.7}} & \cellcolor[HTML]{7BC168}{7.8}           & \cellcolor[HTML]{7DC168}{7.9}          & \cellcolor[HTML]{79C068}{\textbf{7.3}} \\
            $\textrm{VECM}^{\textrm{r1}, \sigma_t, \rho}_{\textrm{lev}, \textrm{ncp}}$   & \cellcolor[HTML]{FFD547}{-2.2}         & \cellcolor[HTML]{FFDF4E}{-2.6}         & \cellcolor[HTML]{FA8C39}{-43.2}        & \cellcolor[HTML]{C6DA60}{2.4}          & \cellcolor[HTML]{F1E85A}{0.7}          & \cellcolor[HTML]{FFB32F}{-8.0}         & \cellcolor[HTML]{A2CE64}{12.0}          & \cellcolor[HTML]{99CB65}{12.3}          & \cellcolor[HTML]{97CA66}{10.4}          & \cellcolor[HTML]{9DCD65}{6.9}           & \cellcolor[HTML]{DBE15E}{2.4}          & \cellcolor[HTML]{FFD346}{-4.1}         \\
            $\textrm{VECM}^{\textrm{r1}, \sigma, \rho_t}_{\textrm{ncp}, \textrm{log}}$   & \cellcolor[HTML]{BDD761}{3.6}          & \cellcolor[HTML]{91C866}{7.6}          & \cellcolor[HTML]{FFC23A}{-6.4}         & \cellcolor[HTML]{F5EA5A}{0.3}          & \cellcolor[HTML]{FFE955}{-0.2}         & \cellcolor[HTML]{FC9335}{-6.6}         & \cellcolor[HTML]{A6CF64}{8.3}           & \cellcolor[HTML]{AFD263}{10.1}          & \cellcolor[HTML]{E4E45C}{4.4}           & \cellcolor[HTML]{88C567}{7.8}           & \cellcolor[HTML]{91C866}{\textbf{8.9}} & \cellcolor[HTML]{CFDD5F}{6.5}          \\
            $\textrm{VECM}^{\textrm{r2}, \sigma_t, \rho}_{}$                             & \cellcolor[HTML]{FFE150}{-0.9}         & \cellcolor[HTML]{FFE754}{-0.9}         & \cellcolor[HTML]{FD9A31}{-26.8}        & \cellcolor[HTML]{D4DF5E}{2.1}          & \cellcolor[HTML]{C5DA60}{2.6}          & \cellcolor[HTML]{FD9931}{-12.2}        & \cellcolor[HTML]{B0D363}{11.8}          & \cellcolor[HTML]{A5CF64}{12.8}          & \cellcolor[HTML]{C3D961}{8.3}           & \cellcolor[HTML]{8DC767}{8.3}           & \cellcolor[HTML]{EEE75B}{1.0}          & \cellcolor[HTML]{FFD648}{-4.0}         \\
            $\textrm{VECM}^{\textrm{r2}, \sigma, \rho_t}_{\textrm{log}}$                 & \cellcolor[HTML]{BBD662}{3.7}          & \cellcolor[HTML]{A2CE64}{7.1}          & \cellcolor[HTML]{FFB934}{-11.6}        & \cellcolor[HTML]{EAE65B}{0.8}          & \cellcolor[HTML]{FFDE4E}{-1.0}         & \cellcolor[HTML]{FC9534}{-11.6}        & \cellcolor[HTML]{9CCC65}{8.8}           & \cellcolor[HTML]{9ECD65}{11.8}          & \cellcolor[HTML]{B4D463}{8.4}           & \cellcolor[HTML]{83C468}{7.5}           & \cellcolor[HTML]{89C667}{8.2}          & \cellcolor[HTML]{C8DB60}{4.8}          \\
            $\textrm{VECM}^{\textrm{r3}, \sigma_t, \rho_t}_{\textrm{lev}, \textrm{ncp}}$ & \cellcolor[HTML]{F1E85A}{0.7}          & \cellcolor[HTML]{DCE15D}{2.7}          & \cellcolor[HTML]{FFA727}{-19.0}        & \cellcolor[HTML]{E3E45D}{1.3}          & \cellcolor[HTML]{DAE15E}{1.6}          & \cellcolor[HTML]{FFB833}{-6.4}         & \cellcolor[HTML]{ADD263}{\textbf{13.4}} & \cellcolor[HTML]{ABD163}{\textbf{13.5}} & \cellcolor[HTML]{A6CF64}{9.4}           & \cellcolor[HTML]{98CB65}{6.1}           & \cellcolor[HTML]{D8E05E}{1.7}          & \cellcolor[HTML]{FFC63D}{-4.1}         \\
            $\textrm{VECM}^{\textrm{r3}, \sigma_t, \rho}_{\textrm{log}}$                 & \cellcolor[HTML]{C4D961}{3.1}          & \cellcolor[HTML]{9BCC65}{8.6}          & \cellcolor[HTML]{FC9633}{-22.2}        & \cellcolor[HTML]{D0DD5F}{1.8}          & \cellcolor[HTML]{E5E55C}{1.3}          & \cellcolor[HTML]{FFA428}{-10.3}        & \cellcolor[HTML]{BAD662}{6.3}           & \cellcolor[HTML]{D1DE5F}{7.1}           & \cellcolor[HTML]{FFC73E}{-17.9}         & \cellcolor[HTML]{8EC766}{8.7}           & \cellcolor[HTML]{A0CD65}{7.2}          & \cellcolor[HTML]{FFE754}{-1.2}         \\
            $\textrm{VECM}^{\textrm{r4}, \sigma_t, \rho}_{\textrm{lev}}$                 & \cellcolor[HTML]{FFD346}{-2.8}         & \cellcolor[HTML]{FFD145}{-5.3}         & \cellcolor[HTML]{F67645}{-52.7}        & \cellcolor[HTML]{E6E55C}{1.1}          & \cellcolor[HTML]{FFCE42}{-2.6}         & \cellcolor[HTML]{FFA329}{-11.5}        & \cellcolor[HTML]{AED263}{11.2}          & \cellcolor[HTML]{B6D562}{9.8}           & \cellcolor[HTML]{FFCF43}{-7.0}          & \cellcolor[HTML]{A7D064}{6.0}           & \cellcolor[HTML]{FDEC58}{0.2}          & \cellcolor[HTML]{FFC83E}{-7.4}         \\
            $\textrm{VECM}^{\textrm{r4}, \sigma, \rho_t}_{\textrm{ncp}, \textrm{log}}$   & \cellcolor[HTML]{D8E05E}{2.3}          & \cellcolor[HTML]{B7D562}{6.5}          & \cellcolor[HTML]{FFA727}{-19.6}        & \cellcolor[HTML]{E0E35D}{1.3}          & \cellcolor[HTML]{FFD346}{-2.6}         & \cellcolor[HTML]{F46D48}{-23.2}        & \cellcolor[HTML]{B9D662}{6.9}           & \cellcolor[HTML]{BDD761}{9.6}           & \cellcolor[HTML]{FFEB56}{-0.8}          & \cellcolor[HTML]{8EC766}{8.2}           & \cellcolor[HTML]{97CA65}{8.9}          & \cellcolor[HTML]{E8E55C}{3.6}          \\
            $\textrm{ETS}^{\sigma}$                                                      & \cellcolor[HTML]{FCEC58}{0.2}          & \cellcolor[HTML]{8DC766}{6.8}          & \cellcolor[HTML]{A4CF64}{5.7}          & \cellcolor[HTML]{C4D961}{1.1}          & \cellcolor[HTML]{B1D363}{0.9}          & \cellcolor[HTML]{FFDF4F}{-0.2}         & \cellcolor[HTML]{99CB65}{10.9}          & \cellcolor[HTML]{98CB65}{11.3}          & \cellcolor[HTML]{94C966}{10.9}          & \cellcolor[HTML]{B7D562}{7.5}           & \cellcolor[HTML]{BCD762}{6.7}          & \cellcolor[HTML]{C7DA60}{5.6}          \\
            $\textrm{ETS}^{\sigma}_{\textrm{log}}$                                       & \cellcolor[HTML]{EFE85B}{1.0}          & \cellcolor[HTML]{80C268}{8.6}          & \cellcolor[HTML]{A0CE64}{\textbf{8.0}} & \cellcolor[HTML]{FCEC58}{0.1}          & \cellcolor[HTML]{E9E65C}{0.7}          & \cellcolor[HTML]{FFDA4B}{-0.6}         & \cellcolor[HTML]{B8D562}{8.9}           & \cellcolor[HTML]{BBD762}{9.4}           & \cellcolor[HTML]{C9DB60}{7.1}           & \cellcolor[HTML]{C1D861}{7.3}           & \cellcolor[HTML]{B7D562}{7.8}          & \cellcolor[HTML]{C8DB60}{6.7}          \\
            $\textrm{VES}^{\sigma}$                                                      & \cellcolor[HTML]{F1634B}{-38.5}        & \cellcolor[HTML]{FFBE38}{-6.4}         & \cellcolor[HTML]{FFBC36}{-5.4}         & \cellcolor[HTML]{F05D4D}{-33.3}        & \cellcolor[HTML]{FFB934}{-6.1}         & \cellcolor[HTML]{FFC43B}{-2.4}         & \cellcolor[HTML]{FE9E2D}{-26.6}         & \cellcolor[HTML]{FFDE4E}{-2.6}          & \cellcolor[HTML]{D7E05E}{3.6}           & \cellcolor[HTML]{F57146}{-37.5}         & \cellcolor[HTML]{FFCB40}{-5.5}         & \cellcolor[HTML]{CBDC60}{4.7}          \\
            $\textrm{VES}^{\sigma}_{\textrm{log}}$                                       & \cellcolor[HTML]{F88140}{-32.4}        & \cellcolor[HTML]{E3E45C}{2.8}          & \cellcolor[HTML]{E8E55C}{1.8}          & \cellcolor[HTML]{F2654B}{-30.4}        & \cellcolor[HTML]{FFB632}{-6.2}         & \cellcolor[HTML]{FFB934}{-3.2}         & \cellcolor[HTML]{FFA626}{-22.0}         & \cellcolor[HTML]{F2E95A}{1.8}           & \cellcolor[HTML]{CFDD5F}{5.4}           & \cellcolor[HTML]{FC9733}{-27.0}         & \cellcolor[HTML]{ECE75B}{2.3}          & \cellcolor[HTML]{AAD164}{6.4}          \\
            \bottomrule
            \multicolumn{13}{l}{\rule{0pt}{1em}Coloring w.r.t. test statistic: \colorbox[HTML]{66BA6A}{<-5} \colorbox[HTML]{7CC168}{-4} \colorbox[HTML]{91C866}{-3} \colorbox[HTML]{B0D363}{-2} \colorbox[HTML]{D8E05E}{-1} \colorbox[HTML]{FFED58}{0} \colorbox[HTML]{FFD145}{1} \colorbox[HTML]{FFB531}{2} \colorbox[HTML]{FC9733}{3} \colorbox[HTML]{F67744}{4} \colorbox[HTML]{EE5250}{>5} }                                                                                                                                                                                                         \\
        \end{tabular}}
\end{table}

The results in Table \ref{tab:crps} align with those of the ES evaluation: \AAr{There is no uniformly superior model in general.  The RW and $\text{VECM}^{r0}$ models seem to perform quite well also in terms of modeling the marginal distributions: in levels for oil and gas prices, in logarithms for EUA prices. For natural gas prices, the levels-VECM models outperform the rest at short horizons, and accounting for leverage effects in the GARCH-part of the model is beneficial.}

Table \ref{tab:rmse} presents the \SP{mean} forecast performance in terms of the RMSE.
\AAr{Considering that no model significantly outperforms the level RW models we find evidence supporting the efficient markets hypothesis.} All RW models perform similarly as distributional modeling has no influence on the mean forecast. Rather, the differences emerge from random simulation fluctuations. \AAr{The models in logarithms, however, seem to perform significantly worse than the level ones, in particular for the EUA prices. This may be attributed to the few extreme trajectories that required capping, as discussed above. }
\begin{table}[!h]
    \caption{Improvement in RMSE score of selected models relative to $\textrm{RW}^{\sigma, \rho}_{}$ in \% (higher = better). Colored according to the test statistic of a DM-Test comparing to $\textrm{RW}^{\sigma, \rho}_{}$  (greener means lower test statistic i.e., better performance compared to $\textrm{RW}^{\sigma, \rho}_{}$). \label{tab:rmse}}
    \centering
    \resizebox{\textwidth}{!}{
        \begin{tabular}[t]{l>{}r>{}r>{}r>{}r>{}r>{}r>{}r>{}r>{}r>{}r>{}r>{}r}
            \toprule
            \multicolumn{1}{c}{ }                                                        & \multicolumn{3}{c}{EUA}                & \multicolumn{3}{c}{Oil}                & \multicolumn{3}{c}{NGas}               & \multicolumn{3}{c}{Coal}                                                                                                                                                                                                                                                                                                                                                        \\
            \cmidrule(l{3pt}r{3pt}){2-4} \cmidrule(l{3pt}r{3pt}){5-7} \cmidrule(l{3pt}r{3pt}){8-10} \cmidrule(l{3pt}r{3pt}){11-13}
            Model                                                                        & H1                                     & H5                                     & H30                                    & H1                                     & H5                                     & H30                                     & H1                                     & H5                                     & H30                                    & H1                                     & H5                                     & H30                                    \\
            \midrule
            $\textrm{RW}^{\sigma, \rho}_{}$                                              & \cellcolor[HTML]{BDBDBD}{\textbf{0.9}} & \cellcolor[HTML]{BDBDBD}{2.0}          & \cellcolor[HTML]{BDBDBD}{5.0}          & \cellcolor[HTML]{BDBDBD}{2.9}          & \cellcolor[HTML]{BDBDBD}{6.4}          & \cellcolor[HTML]{BDBDBD}{\textbf{16.7}} & \cellcolor[HTML]{BDBDBD}{17.8}         & \cellcolor[HTML]{BDBDBD}{42.8}         & \cellcolor[HTML]{BDBDBD}{85.4}         & \cellcolor[HTML]{BDBDBD}{0.9}          & \cellcolor[HTML]{BDBDBD}{2.9}          & \cellcolor[HTML]{BDBDBD}{7.0}          \\
            $\textrm{RW}^{\sigma_t, \rho_t}_{}$                                          & \cellcolor[HTML]{FFE653}{-0.1}         & \cellcolor[HTML]{FFE150}{-0.1}         & \cellcolor[HTML]{B4D462}{0.7}          & \cellcolor[HTML]{FBEC59}{0.0}          & \cellcolor[HTML]{FFC93F}{-0.3}         & \cellcolor[HTML]{FFDF4E}{-0.1}          & \cellcolor[HTML]{FFE14F}{-0.2}         & \cellcolor[HTML]{CADB60}{0.3}          & \cellcolor[HTML]{CFDD5F}{1.3}          & \cellcolor[HTML]{FFE754}{-0.2}         & \cellcolor[HTML]{FFED58}{0.0}          & \cellcolor[HTML]{FFCE42}{-1.8}         \\
            $\textrm{RW}^{\sigma, \rho_t}_{\textrm{ncp}, \textrm{log}}$                  & \cellcolor[HTML]{FFD145}{-270.5}       & \cellcolor[HTML]{FFB833}{-154.1}       & \cellcolor[HTML]{FD9C2F}{-139.9}       & \cellcolor[HTML]{C0D861}{0.5}          & \cellcolor[HTML]{FFDD4D}{-0.5}         & \cellcolor[HTML]{FFD547}{-2.9}          & \cellcolor[HTML]{FFD044}{-0.8}         & \cellcolor[HTML]{E5E45C}{0.7}          & \cellcolor[HTML]{FFE250}{-1.6}         & \cellcolor[HTML]{E2E35D}{0.3}          & \cellcolor[HTML]{FFD044}{-31.2}        & \cellcolor[HTML]{FFBB35}{-24.5}        \\
            $\textrm{RW}^{\sigma, \rho}_{\textrm{log}}$                                  & \cellcolor[HTML]{FFCB40}{-705.0}       & \cellcolor[HTML]{FFC93F}{-265.4}       & \cellcolor[HTML]{FFAD2B}{-125.2}       & \cellcolor[HTML]{B0D363}{0.6}          & \cellcolor[HTML]{F5EA5A}{\textbf{0.2}} & \cellcolor[HTML]{FFE754}{-0.2}          & \cellcolor[HTML]{FFDE4D}{-0.4}         & \cellcolor[HTML]{FBEC59}{0.1}          & \cellcolor[HTML]{FFE251}{-1.6}         & \cellcolor[HTML]{FFC13A}{-0.9}         & \cellcolor[HTML]{FFE250}{-0.3}         & \cellcolor[HTML]{FFC73D}{-8.3}         \\
            $\textrm{VECM}^{\textrm{r0}, \sigma_t, \rho_t}_{\textrm{lev}, \textrm{ncp}}$ & \cellcolor[HTML]{FFDA4B}{-0.9}         & \cellcolor[HTML]{F1E85A}{0.2}          & \cellcolor[HTML]{C6DA60}{0.5}          & \cellcolor[HTML]{E4E45C}{0.5}          & \cellcolor[HTML]{F0E85A}{0.2}          & \cellcolor[HTML]{FFEB56}{0.0}           & \cellcolor[HTML]{FFE754}{-0.4}         & \cellcolor[HTML]{CBDC60}{0.7}          & \cellcolor[HTML]{EFE85B}{0.2}          & \cellcolor[HTML]{E0E35D}{1.4}          & \cellcolor[HTML]{F1E85A}{0.1}          & \cellcolor[HTML]{EAE65B}{0.2}          \\
            $\textrm{VECM}^{\textrm{r0}, \sigma, \rho_t}_{\textrm{log}}$                 & \cellcolor[HTML]{FFD145}{-271.5}       & \cellcolor[HTML]{FFB934}{-191.3}       & \cellcolor[HTML]{FE9F2D}{-114.3}       & \cellcolor[HTML]{CBDC60}{\textbf{1.7}} & \cellcolor[HTML]{FFD145}{-12.3}        & \cellcolor[HTML]{FFC43C}{-3.6}          & \cellcolor[HTML]{FFDF4E}{-0.6}         & \cellcolor[HTML]{E1E35D}{1.6}          & \cellcolor[HTML]{FFDA4B}{-4.1}         & \cellcolor[HTML]{FFEB57}{0.0}          & \cellcolor[HTML]{FFD447}{-0.8}         & \cellcolor[HTML]{FFDD4D}{-6.7}         \\
            $\textrm{VECM}^{\textrm{r1}, \sigma_t, \rho}_{\textrm{lev}, \textrm{ncp}}$   & \cellcolor[HTML]{FFBF38}{-7.7}         & \cellcolor[HTML]{FFD547}{-6.6}         & \cellcolor[HTML]{FFAF2C}{-64.7}        & \cellcolor[HTML]{D2DE5F}{1.2}          & \cellcolor[HTML]{FFD84A}{-2.2}         & \cellcolor[HTML]{FFC139}{-8.4}          & \cellcolor[HTML]{FFC43C}{-3.8}         & \cellcolor[HTML]{FFDF4F}{-1.2}         & \cellcolor[HTML]{FFDD4D}{-5.3}         & \cellcolor[HTML]{DAE15E}{\textbf{1.4}} & \cellcolor[HTML]{FFD849}{-2.0}         & \cellcolor[HTML]{FFE955}{-0.6}         \\
            $\textrm{VECM}^{\textrm{r1}, \sigma, \rho_t}_{\textrm{ncp}, \textrm{log}}$   & \cellcolor[HTML]{FFC13A}{-2.7}         & \cellcolor[HTML]{FFD144}{-86.5}        & \cellcolor[HTML]{FFB430}{-108.7}       & \cellcolor[HTML]{F5EA5A}{0.3}          & \cellcolor[HTML]{FFC83E}{-1.3}         & \cellcolor[HTML]{FD9B2F}{-6.5}          & \cellcolor[HTML]{FFCF43}{-3.0}         & \cellcolor[HTML]{FFEA56}{-0.5}         & \cellcolor[HTML]{FFC53C}{-21.1}        & \cellcolor[HTML]{F7EA59}{0.4}          & \cellcolor[HTML]{FFE351}{-1.1}         & \cellcolor[HTML]{FFC73D}{-24.7}        \\
            $\textrm{VECM}^{\textrm{r2}, \sigma_t, \rho}_{}$                             & \cellcolor[HTML]{FFBD36}{-5.0}         & \cellcolor[HTML]{FFE14F}{-2.3}         & \cellcolor[HTML]{FFB02D}{-40.6}        & \cellcolor[HTML]{D2DE5F}{1.2}          & \cellcolor[HTML]{FCEC58}{0.2}          & \cellcolor[HTML]{FFAE2C}{-33.7}         & \cellcolor[HTML]{FCEC59}{0.2}          & \cellcolor[HTML]{DDE25D}{\textbf{3.9}} & \cellcolor[HTML]{F2E95A}{\textbf{1.6}} & \cellcolor[HTML]{E9E65C}{1.0}          & \cellcolor[HTML]{F1E85A}{\textbf{0.5}} & \cellcolor[HTML]{FFE754}{-1.4}         \\
            $\textrm{VECM}^{\textrm{r2}, \sigma, \rho_t}_{\textrm{log}}$                 & \cellcolor[HTML]{FFBA35}{-3.8}         & \cellcolor[HTML]{FFC63D}{-117.4}       & \cellcolor[HTML]{FEA12B}{-141.4}       & \cellcolor[HTML]{CDDC60}{1.6}          & \cellcolor[HTML]{FFC93F}{-11.8}        & \cellcolor[HTML]{FC9534}{-18.8}         & \cellcolor[HTML]{FFCB40}{-2.1}         & \cellcolor[HTML]{E3E45C}{1.8}          & \cellcolor[HTML]{FFD346}{-10.4}        & \cellcolor[HTML]{FCEC59}{0.1}          & \cellcolor[HTML]{FFD145}{-44.5}        & \cellcolor[HTML]{FFC93F}{-12.9}        \\
            $\textrm{VECM}^{\textrm{r3}, \sigma_t, \rho_t}_{\textrm{lev}, \textrm{ncp}}$ & \cellcolor[HTML]{FFBE37}{-4.3}         & \cellcolor[HTML]{FFEC57}{-0.2}         & \cellcolor[HTML]{FFA428}{-29.8}        & \cellcolor[HTML]{FFE855}{-0.2}         & \cellcolor[HTML]{FFDC4C}{-1.1}         & \cellcolor[HTML]{FFBD37}{-5.8}          & \cellcolor[HTML]{FAEC59}{\textbf{0.2}} & \cellcolor[HTML]{BDD761}{2.0}          & \cellcolor[HTML]{FFED58}{-0.1}         & \cellcolor[HTML]{DEE25D}{1.0}          & \cellcolor[HTML]{FFEA56}{-0.1}         & \cellcolor[HTML]{F3E95A}{0.9}          \\
            $\textrm{VECM}^{\textrm{r3}, \sigma_t, \rho}_{\textrm{log}}$                 & \cellcolor[HTML]{FFB934}{-4.5}         & \cellcolor[HTML]{FFC53C}{-140.2}       & \cellcolor[HTML]{F9873D}{-877.4}       & \cellcolor[HTML]{C8DB60}{1.5}          & \cellcolor[HTML]{FFE552}{-0.8}         & \cellcolor[HTML]{FC9534}{-19.3}         & \cellcolor[HTML]{FFCB41}{-5.7}         & \cellcolor[HTML]{FFDA4B}{-6.1}         & \cellcolor[HTML]{FFB732}{-74.6}        & \cellcolor[HTML]{F4E95A}{0.7}          & \cellcolor[HTML]{FFDE4D}{-1.6}         & \cellcolor[HTML]{FFC73D}{-15.5}        \\
            $\textrm{VECM}^{\textrm{r4}, \sigma_t, \rho}_{\textrm{lev}}$                 & \cellcolor[HTML]{FFBE37}{-8.8}         & \cellcolor[HTML]{FFC53C}{-12.1}        & \cellcolor[HTML]{FE9E2D}{-79.5}        & \cellcolor[HTML]{FFDC4C}{-0.7}         & \cellcolor[HTML]{FFB833}{-4.5}         & \cellcolor[HTML]{FFAC2B}{-10.0}         & \cellcolor[HTML]{FFD749}{-1.6}         & \cellcolor[HTML]{FFDD4D}{-2.1}         & \cellcolor[HTML]{FFA827}{-32.3}        & \cellcolor[HTML]{ECE75B}{0.8}          & \cellcolor[HTML]{FFE351}{-0.7}         & \cellcolor[HTML]{FFE452}{-2.3}         \\
            $\textrm{VECM}^{\textrm{r4}, \sigma, \rho_t}_{\textrm{ncp}, \textrm{log}}$   & \cellcolor[HTML]{FFC139}{-4.1}         & \cellcolor[HTML]{FFD94A}{-3.3}         & \cellcolor[HTML]{FD9A30}{-40.9}        & \cellcolor[HTML]{DCE15D}{1.6}          & \cellcolor[HTML]{FFCA40}{-4.8}         & \cellcolor[HTML]{F88040}{-25.1}         & \cellcolor[HTML]{FFCB40}{-5.0}         & \cellcolor[HTML]{FFE351}{-2.5}         & \cellcolor[HTML]{FFC139}{-31.7}        & \cellcolor[HTML]{F2E95A}{0.7}          & \cellcolor[HTML]{F8EB59}{0.4}          & \cellcolor[HTML]{EFE85B}{\textbf{3.0}} \\
            $\textrm{ETS}^{\sigma}$                                                      & \cellcolor[HTML]{FFE452}{-0.3}         & \cellcolor[HTML]{F0E85B}{0.3}          & \cellcolor[HTML]{D8E05E}{1.6}          & \cellcolor[HTML]{D0DD5F}{0.7}          & \cellcolor[HTML]{F4E95A}{0.1}          & \cellcolor[HTML]{FFDD4D}{-0.1}          & \cellcolor[HTML]{F0E85A}{0.1}          & \cellcolor[HTML]{FFD648}{-0.1}         & \cellcolor[HTML]{E5E45C}{0.2}          & \cellcolor[HTML]{FFD849}{-2.4}         & \cellcolor[HTML]{FFC63D}{-3.9}         & \cellcolor[HTML]{E6E55C}{2.5}          \\
            $\textrm{ETS}^{\sigma}_{\textrm{log}}$                                       & \cellcolor[HTML]{FFCE43}{-1.0}         & \cellcolor[HTML]{F5EA5A}{\textbf{0.4}} & \cellcolor[HTML]{ECE75B}{\textbf{1.6}} & \cellcolor[HTML]{C7DA60}{0.9}          & \cellcolor[HTML]{FFED58}{0.0}          & \cellcolor[HTML]{FFEA56}{-0.1}          & \cellcolor[HTML]{FFD447}{-1.9}         & \cellcolor[HTML]{FFE150}{-1.9}         & \cellcolor[HTML]{FFCE42}{-13.9}        & \cellcolor[HTML]{FFE250}{-0.3}         & \cellcolor[HTML]{FFCD42}{-3.6}         & \cellcolor[HTML]{FFE452}{-1.8}         \\
            $\textrm{VES}^{\sigma}$                                                      & \cellcolor[HTML]{FFA828}{-37.4}        & \cellcolor[HTML]{FFBC36}{-8.9}         & \cellcolor[HTML]{FFBC36}{-6.0}         & \cellcolor[HTML]{FD9931}{-27.9}        & \cellcolor[HTML]{FFC53C}{-7.4}         & \cellcolor[HTML]{FFB430}{-2.8}          & \cellcolor[HTML]{FFAB2A}{-27.2}        & \cellcolor[HTML]{FFB732}{-9.5}         & \cellcolor[HTML]{FFDD4D}{-2.4}         & \cellcolor[HTML]{FFC038}{-41.7}        & \cellcolor[HTML]{FFDA4B}{-1.2}         & \cellcolor[HTML]{D4DF5F}{1.6}          \\
            $\textrm{VES}^{\sigma}_{\textrm{log}}$                                       & \cellcolor[HTML]{FFAA29}{-37.6}        & \cellcolor[HTML]{FFBD37}{-9.2}         & \cellcolor[HTML]{FFBA35}{-7.8}         & \cellcolor[HTML]{FD9A30}{-26.8}        & \cellcolor[HTML]{FFC63D}{-7.3}         & \cellcolor[HTML]{FFBE37}{-3.0}          & \cellcolor[HTML]{FFAC2A}{-27.0}        & \cellcolor[HTML]{FFC33B}{-6.8}         & \cellcolor[HTML]{FFD144}{-3.5}         & \cellcolor[HTML]{FFC139}{-41.2}        & \cellcolor[HTML]{FFC43C}{-2.2}         & \cellcolor[HTML]{FFE855}{-0.3}         \\
            \bottomrule
            \multicolumn{13}{l}{\rule{0pt}{1em}Coloring w.r.t. test statistic: \colorbox[HTML]{66BA6A}{<-5} \colorbox[HTML]{7CC168}{-4} \colorbox[HTML]{91C866}{-3} \colorbox[HTML]{B0D363}{-2} \colorbox[HTML]{D8E05E}{-1} \colorbox[HTML]{FFED58}{0} \colorbox[HTML]{FFD145}{1} \colorbox[HTML]{FFB531}{2} \colorbox[HTML]{FC9733}{3} \colorbox[HTML]{F67744}{4} \colorbox[HTML]{EE5250}{>5} }                                                                                                                                                                                                      \\
        \end{tabular}}
\end{table}

\begin{figure}[!h]
    \centering
    \resizebox{\textwidth}{!}{%
        \fbox{\input{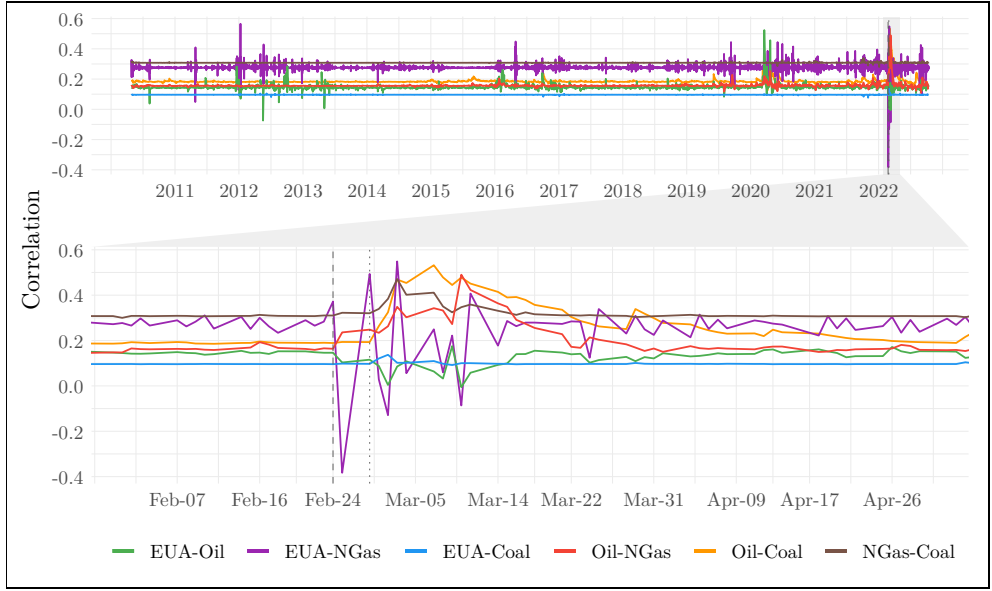}}
    }
    \caption{\JB{Temporal evolution of the linear dependence as provided by time varying dependence parameter modeling in the $\textrm{VECM}^{\textrm{r0}, \sigma, \rho_t}_{\textrm{log}}$ model. The dashed line marks the Russian invasion of Ukraine (2022-02-24). The dotted line marks the freeze of foreign reserves of the central bank of Russia (2022-02-28).}}
    \label{r99_rho}
\end{figure}

\let\pgfimageWithoutPath\pgfimage
\renewcommand{\pgfimage}[2][]{\pgfimageWithoutPath[#1]{anc/#2}}
\AAr{We next evaluate the  temporal evolution of the copula-implied pairwise correlations of the best-performing (in terms of ES) model $\text{VECM}^{r0, \sigma, \rho_t}_{\text{log}}$ presented in Figure \ref{r99_rho}.} The correlations between the prices of natural gas and coal, and EUA and coal are approximately constant over time.
\JB{The correlations of oil with coal and gas fluctuate around a level of 0.2 featuring a temporary increase of up to 0.5 in the emerging Russian invasion.}
\SP{The} most volatile correlation is that between EUA and natural gas prices. \SP{It fluctuates strongly around a level of approximately 0.3.} Immediately after the Russian invasion of Ukraine, the correlation between EUA and gas prices changes dramatically, dropping from a level of about \SP{0.3} to nearly \SP{-0.4.}
\SP{By mid-March 2022 the relation has stabilized.}

\begin{figure}[!h]
    \centering
    \resizebox{\textwidth}{!}{%
        \fbox{\input{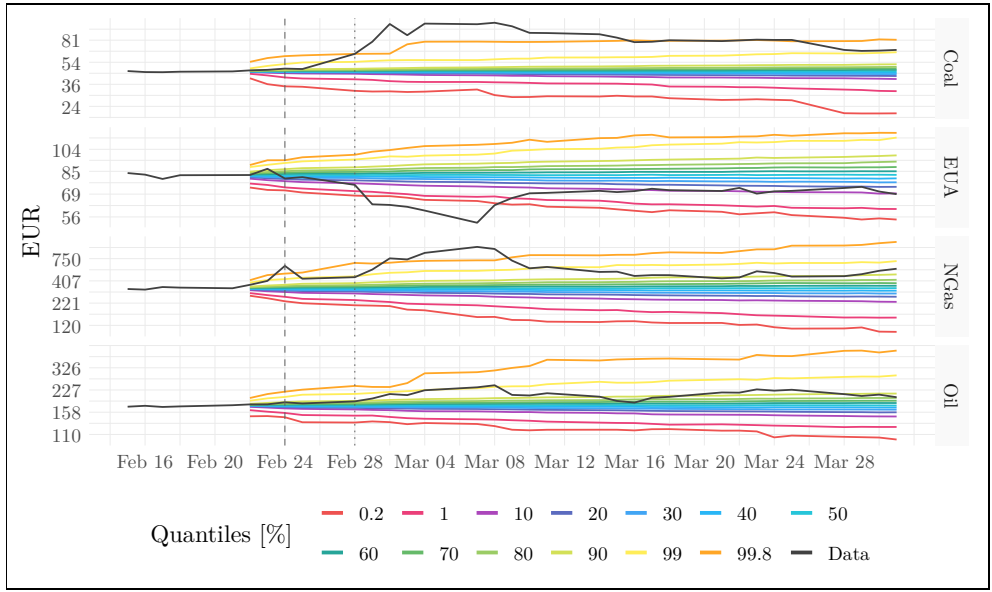}}
    }
    \caption{\JB{Predictive quantiles in \% extracted from 2048 trajectories predicted by the the $\textrm{VECM}^{\textrm{r0}, \sigma, \rho_t}_{\textrm{log}}$ model starting 2022-02-21. The plot depicts inflation-adjusted and emission-normalized data. The dashed line marks the Russian invasion of Ukraine (2022-02-24). The dotted line marks the freeze of foreign reserves of the central bank of Russia (2022-02-28).}}
    \label{fig_quant}
\end{figure}

\AAr{Finally, Figure \ref{fig_quant} displays the predictive quantiles in percent extracted from the 2048 trajectories of the $\textrm{VECM}^{\textrm{r0}, \sigma, \rho_t}_{\textrm{log}}$ model starting right before the Russian invasion in Ukraine. Evidently, the prices' development in the first few days thereafter does not seem too extreme with the exception of the gas price peak on February 24 itself. However, coal, EUA, and gas prices drift well beyond the projected $99.8^{\text{th}}$ percentile  in the three weeks after the subsequent freeze of the foreign reserves of Russia's central bank, suggesting that this reaction had been largely unanticipated by the markets.  }

\section{Conclusion}\label{sec_conclusion}
In this paper, we propose a VECM-Copula-GARCH model to jointly model EUA, oil, coal, and natural gas prices with the aim of producing short-term probabilistic forecasts. \AAr{The price series are in real terms and standardized to reflect one tonne of CO$_2$ emissions for proper comparison.} Allowing time-varying dependence parameters in the copula and leverage effects in the model of the conditional heteroscedasticity allows us not only to utilize important data characteristics to improve the forecasts, but also to analyze the temporal evolution of the estimated parameters, providing insight into the linkages between EUA and fossil fuels markets. An extensive rolling-window forecasting study compares the proposed model with relevant \AAr{benchmark models for both the data in levels and after log-transformation.

The best-performing model is $\textrm{VECM}^{\textrm{r0}, \sigma, \rho_t}_{\textrm{log}}$, i.e. a VAR with constant conditional variance and time-varying dependence parameters estimated on log-transformed data. In general, stabilizing the variance by the log-transformation and employing multivariate random walk models is more beneficial for the overall forecasting performance than accounting for long-term relationships between the variables in levels. The latter can be preferable for forecasts at short horizons, but only in conjunction with modeling the time variation in the conditional volatility, which accounts for the greatest share of improvement in the forecasts. The improvement contributed by time-varying copula dependence, leverage effects in the GARCH-part of the model and marginal distribution parameters is less pronounced.}

\bibliography{bibliography}

\end{document}